\documentclass[aps,pra]{revtex4-2}
\usepackage{amsfonts}
\usepackage{amsmath}
\usepackage{amsthm}
\usepackage{amssymb}
\usepackage{graphicx}
\usepackage{subfigure}

\begin{document}

\title{New findings for the old problem: Exact solutions for domain walls in
coupled real Ginzburg-Landau equations}
\author{Boris A. Malomed$^{1.2}$}
\affiliation{$^{1}$Department of Physical Electronics, School of Electrical Engineering,
Faculty of Engineering, and Center for Light-Matter Interaction, Tel Aviv
University, P.O. Box 39040 Tel Aviv, Israel\\
$^{2}$Instituto de Alta Investigaci\'{o}n, Universidad de Tarapac\'{a}, Casilla 7D,
Arica, Chile}

\begin{abstract}
This work reports new exact solutions for domain-wall (DW) states produced
by a system of coupled real Ginzburg-Landau (GL) equations which model
patterns in thermal convection, optics, and Bose-Einstein condensates
(BECs). An exact solution for symmetric DW was known for a single value of
the cross-interaction coefficient, $G=3$ (defined so that its
self-interaction counterpart is $1$). Here an exact asymmetric DW is
obtained for the system in which the diffusion term is absent in one
component. It exists for all $G>1$. Also produced is an exact solution for
DW in the symmetric real-GL system which includes linear coupling. In
addition, an effect of a trapping potential on the DW is considered, which
is relevant to the case of BEC. In a system of three GL equations, an exact
solution is obtained for a composite state including a two-component DW and
a localized state in the third component. Bifurcations which create two
lowest composite states are identified too. Lastly, exact solutions are
found for the system of real GL equations for counterpropagating waves,
which represent a sink or source of the waves, as well as for a system of
three equations which includes a standing localized component.
\end{abstract}

\maketitle

\noindent \textbf{Keywords}: Rayleigh-B\'{e}nard convection; pattern
formation; Lyapunov functional; grain boundary; Thomas-Fermi approximation;
linear coupling

\section{Introduction}

Complex Ginzburg-Landau (GL) equations is a well-known class of fundamental
models underlying the theory of pattern formation under the combined action
of linear gain and loss (including diffusion/viscosity), linear wave
dispersion, nonlinear loss, and nonlinear dispersion. In the case of the
cubic nonlinearity, the generic one-dimensional form of this equation for a
complex order parameter, $u(x,t)$, is \cite{AransonKramer,Encycl}
\begin{equation}
\frac{\partial u}{\partial t}=gu+\left( a+ib\right) \frac{\partial ^{2}u}{%
\partial x^{2}}-\left( d+ic\right) |u|^{2}u.  \label{CCGL}
\end{equation}%
Here, positive constants $g,d$, and $a$ represent, severally, the linear
gain, nonlinear loss, and diffusion. Coefficients $b$ and $c$, which may
have any sign, control the linear and nonlinear dispersion, respectively. By
means of obvious rescaling of $t$, $x$, and $u$, one can fix three
coefficients in Eq. (\ref{CCGL}):%
\begin{equation}
g=d=a=1.  \label{11}
\end{equation}

The ubiquity and great variety of the complex GL equations is illustrated by
the title of the well-known review article by Aranson and Kramer \cite%
{AransonKramer}, \textit{The world of the complex Ginzburg-Landau equation}.
These equations are directly derived in settings such as laser cavities,
with $u(x,t)$ being a slowly-varying amplitude of the optical field \cite%
{Arrechi,Rosanov1,Rosanov2,Inc}. In many other areas (hydrodynamics,
plasmas, chemical waves, etc.), underlying systems of basic equations are
more cumbersome, but complex GL equations can be derived as asymptotic ones
governing the evolution of long-scale small-amplitude (but, nevertheless,
essentially nonlinear) excitations \cite{CrossHohenberg,Kramer,Hoyle}.

A particular case of Eq. (\ref{CCGL}) is the real GL equation (in this
context, the name had originally appeared from the phenomenological theory
of superconductivity elaborated by Ginzburg and Landau 70 years ago \cite%
{GiLa}):%
\begin{equation}
\frac{\partial u}{\partial t}=u+\frac{\partial ^{2}u}{\partial x^{2}}%
-|u|^{2}u,  \label{realGL}
\end{equation}%
which is written with respect to normalization (\ref{11}). Actually, order
parameter $u(x,t)$ governed by Eq. (\ref{realGL}) may be a complex function,
while the equation is called \textquotedblleft real" because its
coefficients are real. The real GL equation is well known as a model of
nondispersive nonlinear dissipative media, such as the Rayleigh-B\'{e}nard
(RB) convection in a layer of a fluid heated from below \cite%
{Busse,Cross1982}, and instability of a plane laser-driven evaporation front
\cite{Anisimov}.

Unlike Eq. (\ref{CCGL}) with complex coefficients, real GL equation (\ref%
{realGL}) may be represented in the \textit{gradient form}, $\partial
u/\partial t=-\delta L/\delta u^{\ast }$, where $\delta /\delta u^{\ast }$
stands for the variational (Frech\'{e}) derivative, and
\begin{equation}
L=\int_{-\infty }^{+\infty }\left( -|u|^{2}+\left\vert \frac{\partial u}{%
\partial x}\right\vert ^{2}+\frac{1}{2}|u|^{4}\right) dx  \label{Lyapunov}
\end{equation}%
is the \textit{Lyapunov functional}. A consequence of the gradient
representation is that $L$ may only decrease or stay constant in the course
of the evolution, $dL/dt\leq 0$. This fact strongly simplifies dynamics of
the real GL equation, especially the study of stability of its stationary
solutions.

Equation (\ref{realGL}) gives rise to a family of stationary plane-wave (PW)
solutions,%
\begin{equation}
u(x)=\sqrt{1-k^{2}}\exp \left( ikx\right) ,  \label{CW}
\end{equation}%
where real wavenumber $k$ takes values in the existence band, $-1<k<+1$. In
terms of the RB convection, the PWs represent the simplest nontrivial
patterns in the form of periodic arrays of counter-rotating convective
\textquotedblleft rolls", which appear when the Rayleigh number exceeds its
critical value \cite{Busse,Cross1982}. The PW solutions are stable against
small perturbations in a part of the existence band, which is selected by
the \textit{Eckhaus criterion} \cite{Eckhaus,early}: $-1/\sqrt{3}\leq k\leq
+1/\sqrt{3}$. In the stability subband, the squared amplitude of the PW
solution, $A^{2}(k)$, must exceed $2/3$ of its maximum value, $A_{\max
}^{2}\equiv 1$, which corresponds to $k=0$:%
\begin{equation}
A^{2}(k)\equiv 1-k^{2}\geq 2/3.  \label{2/3}
\end{equation}%
The density of the Lyapunov functional (\ref{Lyapunov}) of the PW solutions,
$\mathcal{L}=-A^{4}(k)/2$, takes values $\mathcal{L}_{\min }\equiv -1/2\leq
\mathcal{L}\leq (4/9)\mathcal{L}_{\min }$, as $k^{2}$ varies from $0$ to $%
1/3 $ in the stability interval (\ref{2/3}). The presence of the interval of
values of $k$ which give rise to stable roll patterns puts forward the
problem of the \textit{wavenumber selection}, which was addressed in various
settings \cite{Eshel,Misha,Hecke,Scheel}.

In fact, the rolls are quasi-one-dimensional patterns, as the surface of the
convection layer is two-dimensional. This fact suggests a possibility of the
existence of patterns with linear defects in the form of domain walls (DWs),
alias grain boundaries, separating half-infinite areas filled by PWs with
wave vectors $\mathbf{k}_{1,2}$ with different orientations but equal
lengths, $k_{1,2}=1$. Such defects may be naturally formed by the
Kibble-Zurek mechanism \cite{KZ,KZ-Laroze}, when the switch of the Rayleigh
number of the fluid layer heated from below to a supercritical value, at
which the convection instability sets in, occurs at two separated spots.
They become sources of rolls with independently chosen orientations.
Collision between arrays of rolls with different orientation will naturally
give rise to an interface in the form of the DW. These structures in the RB
convection were predicted theoretically \cite%
{Cross1982,Manneville,Trib-DW,Iooss} and observed in experiments, both as
DWs proper and more complex defects, formed by intersecting DWs \cite%
{Steinberg}. Actually, the existence of the DW is a consequence of the
effective immiscibility of the PW modes \cite{Mineev,immiscible} which are
separated by the wall.

It is relevant to mention that grain boundaries occur, in a great variety of
realizations, as fundamental objects in condensed-matter physics \cite%
{grain1,grain2,grain4,grain5,grain6,grain3}. Although the nature of such
objects is different from that in the RB convection and other nonlinear
dissipative media, the phenomenology of the grain boundaries has many common
features in all physical settings where they appear.

DW states were constructed in Ref. \cite{Trib-DW} as solutions of two
coupled real GL equations for amplitudes $u_{1}$ and $u_{2}$ of PWs
connected by the DW, see Eqs. (\ref{1}) and (\ref{2}) below. At the level of
stationary solutions, the same coupled real equations predict DWs in optics,
as boundaries between spatial or temporal domains occupied by PWs
representing different polarizations or different wavelengths of light \cite%
{optical-DW,optical-DW2}. Further, these equations coincide with the
stationary version of the Gross-Pitaevskii (GP) equations which produce DW
states in binary Bose-Einstein condensates (BECs) composed of immiscible
components \cite{Poland,BEC-DW}.

In a particular case, a DW solution of the coupled real GL equations was
found in an exact analytical form, see Eq. (\ref{exact1}) below. Although
the exact solution is not a generic one, it is an obviously important
finding, as it provides direct insight into the structure of the respective
states. The objective of the present work is to add several new exact
solutions of the DW type for more general forms of coupled real GL
equations, which exhibit essentially new features. The new solutions are:
(i) an exact DW state in the extremely asymmetric system, in which the
diffusion coefficient vanishes in one equation; (ii) the system including
linear coupling between the components; (iii) a composite state including a
DW in two components and a bright soliton in the additional component added
to the system; (iv) an exact DW-like state of the \textit{source} or \textit{%
sink} types in a system of GL equations for counterpropagating waves, which
is a basic model for the traveling-wave thermal convection in binary fluids
\cite{Cross,Cross2,Kolodner}. It was known that the interplay of
counterpropagating waves could give rise to source and sink modes \cite%
{Pomeau,Coullet,traveling,Burgers}, but exact solutions for them were not
available. Also reported is an exact composite solution of a system of three
equations, in the form of a source or kink formed by two counterpropagating
components, coupled to a localized standing mode in the third component. An
essential fact is that, unlike the particular exact DW solution originally
reported in Ref. \cite{Trib-DW}, which was an isolated one, with no degrees
of freedom, the new solutions reported here appear in \textit{families},
which contain at least one free parameter.

The above-mentioned new exact solutions are presented, respectively, in
Sections II -- V. In addition, Section III addresses the situation relevant
to the realization of the coupled system in BEC, when the GP equations
include a trapping harmonic-oscillator (HO)\ potential. Section IV also
reports exact results for bifurcations which create two lowest
three-component composite states, with an infinitesimal even or odd mode in
the third component, added to the DW. The paper is concluded by Section VI

\section{The DW (domain wall) in the asymmetric system}

The starting point of the analysis leading to the coupled system of GL
equations for slowly-varying amplitudes of $N$ two-dimensional PWs, $%
u_{j}\left( x,y,t\right) $, $j=1,...,N$, with carrier wave vectors $\mathbf{k%
}_{j}$ of the PWs which form a convection pattern, or a similar one in other
physical setups, is the expression for the two-dimensional distribution of
the complex order parameter (e.g., the amplitude of the convective flow):%
\begin{equation}
U(x,y;t)=\sum_{l=1}^{N}u_{j}(x,y;t)\exp \left( i\mathbf{k}_{l}\mathbf{\cdot r%
}\right) ,  \label{U}
\end{equation}%
where \textbf{$r$}$\mathbf{=}\left( x,y\right) $ \cite%
{Busse,Cross1982,Trib-DW}. In the case of $N=2$, the resulting system of
coupled one-dimensional GL equations for the configuration which represents
the DW oriented along axis $x$ is, in the scaled form,%
\begin{eqnarray}
\frac{\partial u_{1}}{\partial t} &=&D_{1}\frac{\partial ^{2}u_{1}}{\partial
x^{2}}+u_{1}\left( 1-\left\vert u_{1}\right\vert ^{2}-G\left\vert
u_{2}\right\vert ^{2}\right) ,  \label{1d/dt} \\
\frac{\partial u_{1}}{\partial t} &=&D_{2}\frac{\partial ^{2}u_{2}}{\partial
x^{2}}+u_{2}\left( 1-\left\vert u_{2}\right\vert ^{2}-G\left\vert
u_{1}\right\vert ^{2}\right) ,  \label{2d/dt}
\end{eqnarray}%
cf. Eq. (\ref{realGL}). Here, $G>0$ is the coefficient of the inter-mode
interaction, while its counterpart for the self-interaction is scaled to be $%
1$, and diffusion coefficients are
\begin{equation}
D_{1.2}\equiv \cos ^{2}\theta _{1,2},  \label{D}
\end{equation}%
where $\theta _{1,2}$ are angles between vectors $\mathbf{k}_{1,2}$ and the $%
x$ axis. Similar to Eq. (\ref{realGL}), this system may be written in the
gradient form, $\partial u_{1,2}/\partial t=-\delta L/\delta u_{1,2}^{\ast }$%
, with the Lyapunov functional which is an extension of expression (\ref%
{Lyapunov}):
\begin{equation}
L_{\mathrm{12}}=\int_{-\infty }^{+\infty }\left[ \sum_{j=1,2}\left(
-|u_{j}|^{2}+\left\vert \frac{\partial u_{j}}{\partial x}\right\vert ^{2}+%
\frac{1}{2}|u_{j}|^{4}\right) +G\left\vert u_{1}\right\vert ^{2}\left\vert
u_{2}\right\vert ^{2}\right] dx.  \label{L2}
\end{equation}%
Real DW solutions, interpolating between uniform PW modes $u_{1}$ and $u_{2}$
at $x\rightarrow -\infty $ and $x\rightarrow +\infty $, respectively,
satisfy the stationary version of Eqs. (\ref{1d/dt}) and (\ref{2d/dt}),%
\begin{eqnarray}
D_{1}\frac{d^{2}u_{1}}{dx^{2}}+u_{1}\left( 1-u_{1}^{2}-Gu_{2}^{2}\right)
&=&0,  \label{1} \\
D_{2}\frac{d^{2}u_{2}}{dx^{2}}+u_{2}\left( 1-u_{2}^{2}-Gu_{1}^{2}\right)
&=&0,  \label{2}
\end{eqnarray}%
and are determined by boundary conditions (b.c.)%
\begin{gather}
u_{1}\left( x\rightarrow -\infty \right) =u_{2}\left( x\rightarrow +\infty
\right) =1,  \notag \\
u_{1}\left( x\rightarrow +\infty \right) =u_{2}\left( x\rightarrow -\infty
\right) =0.  \label{bc}
\end{gather}%
These solutions exists under the above-mentioned immiscibility constraints
which, in the present notation, is $G>1$ (i.e., the inter-component
repulsion is stronger than the intrinsic self-repulsion in each component)
\cite{Mineev}. If DW solutions to Eqs. (\ref{1}) and (\ref{2}) exist, the
Lyapunov functional (\ref{L2}) guarantees their stability (it can be checked
that they correspond to minima of the functional, rather than to a saddle
point).

An essential finding, reported in Ref. \cite{Trib-DW}, is that the symmetric
version of Eqs. (\ref{1}) and (\ref{2}), with $D_{1}=D_{2}\equiv D$,
produces a particular exact solution:%
\begin{equation}
G=3,\left\{
\begin{array}{c}
u_{1}(x) \\
u_{2}(x)%
\end{array}%
\right\} =\frac{1}{2}\left\{
\begin{array}{c}
1-\tanh \left( x/\sqrt{2D}\right) \\
1+\tanh \left( x/\sqrt{2D}\right)%
\end{array}%
\right\} .  \label{exact1}
\end{equation}%
In terms of the BEC realization of Eq. (\ref{1}) and (\ref{2}), the single
value of the interaction coefficient at which this exact solution is
available, $G=3$, can be adjusted by means of the Feshbach-resonance method
for binary condensates \cite{Feshbach,Feshbach2}. In optics, the usual value
is $G=2$ for the copropagation of waves with orthogonal circular
polarizations or different wavelengths, but other values of $G$ can be
adjusted in nonlinear photonic crystals \cite{phot-cryst}.

The first new result, reported here as an essential addition to the
well-elaborated theme of DWs, is the fact that it is also possible to find
an exact analytical solution in the limit case of the extreme asymmetry in
the system of Eqs. (\ref{1}) and (\ref{2}), which corresponds to $D_{2}=0$
and $D_{1}\equiv D>0$, i.e., the DW between two roll families one of which
has the wave vector perpendicular to the $x$ axis, see Eq. (\ref{D}):
\begin{gather}
D\frac{d^{2}u_{1}}{dx^{2}}+u_{1}\left( 1-u_{1}^{2}-Gu_{2}^{2}\right) =0,
\label{D1} \\
u_{2}\left( 1-u_{2}^{2}-Gu_{1}^{2}\right) =0.  \label{D=0}
\end{gather}%
Note that the form of Eq. (\ref{D=0}), in which the second derivative drops
out, corresponds to the well-known Thomas-Fermi (TF) approximation\ in the
BEC\ theory. In the framework of the TF approximation, the kinetic-energy
term in the GP\ equation is neglected, in comparison with ones representing
a trapping potential and the self-repulsive nonlinearity \cite{Pit}. In the
present case, $D_{2}=0$ is not an approximation, but the exact special case
corresponding to $\theta _{2}=90^{\mathrm{o}}$ in Eq. (\ref{D}). As concerns
the application of Eqs. (\ref{1}) and (\ref{2}), as a system of stationary
GP equations, to BEC, with the kinetic-energy coefficients which are $%
D_{1,2}=\hbar ^{2}/\left( 2m_{1,2}\right) $ in physical units, where $%
m_{1,2} $ are atomic masses of the two components of the heteronuclear
binary condensate, Eqs. (\ref{D1}) and (\ref{D=0}) correspond to a \textit{%
semi-TF approximation}, representing a mixture of light (small $m_{1}$) and
heavy (large $m_{2}$) atoms, e.g., a $^{7}$Li--$^{87}$Rb diatomic gas \cite%
{Li-Rb}.

Obviously, Eq. (\ref{D=0}) yields two solutions, \textit{viz}., either $%
u_{2}=0$, or one representing the quasi-TF relation,
\begin{equation}
u_{2}^{2}(x)=1-Gu_{1}^{2}(x).  \label{TF}
\end{equation}%
Equation (\ref{D1}) with $u_{2}=0$ yields the usual dark soliton, while the
substitution of expression (\ref{TF}) in Eq. (\ref{D1}) may produce a
bright-soliton solution. These solutions are matched at a stitch point,
\begin{equation}
x=x_{0}\equiv -\sqrt{\frac{D}{2}}\ln \left( \frac{\sqrt{G}+1}{\sqrt{G}-1}%
\right) ,  \label{x0}
\end{equation}%
which is defined by condition $u_{1}^{2}(x)=1/G$, according to Eq. (\ref{TF}%
). The global solution, which complies with b.c. (\ref{bc}), is%
\begin{equation}
u_{1}(x)=\left\{
\begin{array}{c}
-\tanh \left( x/\sqrt{2D}\right) ,~\mathrm{at}~-\infty <x<x_{0}, \\
\sqrt{\frac{2}{G+1}}\mathrm{sech}\left[ \sqrt{\frac{G-1}{D}}\left( x-\xi
\right) \right] ,\mathrm{at}~x_{0}<x<+\infty ,%
\end{array}%
\right.  \label{r1exact}
\end{equation}%
\begin{equation}
u_{2}(x)=\left\{
\begin{array}{c}
0,~\mathrm{at}~-\infty <x<x_{0}, \\
\sqrt{1-Gu_{1}^{2}(x)},\mathrm{at}~x_{0}<x<+\infty .%
\end{array}%
\right.  \label{r2exact}
\end{equation}%
Finally, the virtual center of the bright-soliton segment of $u_{1}(x)$ is
located at
\begin{equation}
x=\xi \equiv x_{0}-\sqrt{\frac{D}{G-1}}\ln \left( \sqrt{\frac{2G}{G+1}}+%
\sqrt{\frac{G-1}{G+1}}\right)  \label{xi}
\end{equation}%
(actually, exact solution (\ref{r1exact}) includes the \textquotedblleft
tail" of the bright soliton at $x\geq x_{0}$, which does not cover the
central point, $x=\xi $). The distance $x_{0}-\xi $, determined by Eq. (\ref%
{xi}), defines the effective thickness of the strongly asymmetric DW. Note
that, as seen from Eqs. (\ref{x0}) and (\ref{r1exact}), this exact solution
exists under the constraint of $G>1$, which is the above-mentioned
immiscibility condition.

\begin{figure}[tbp]
\begin{center}
\includegraphics[width=0.60\textwidth]{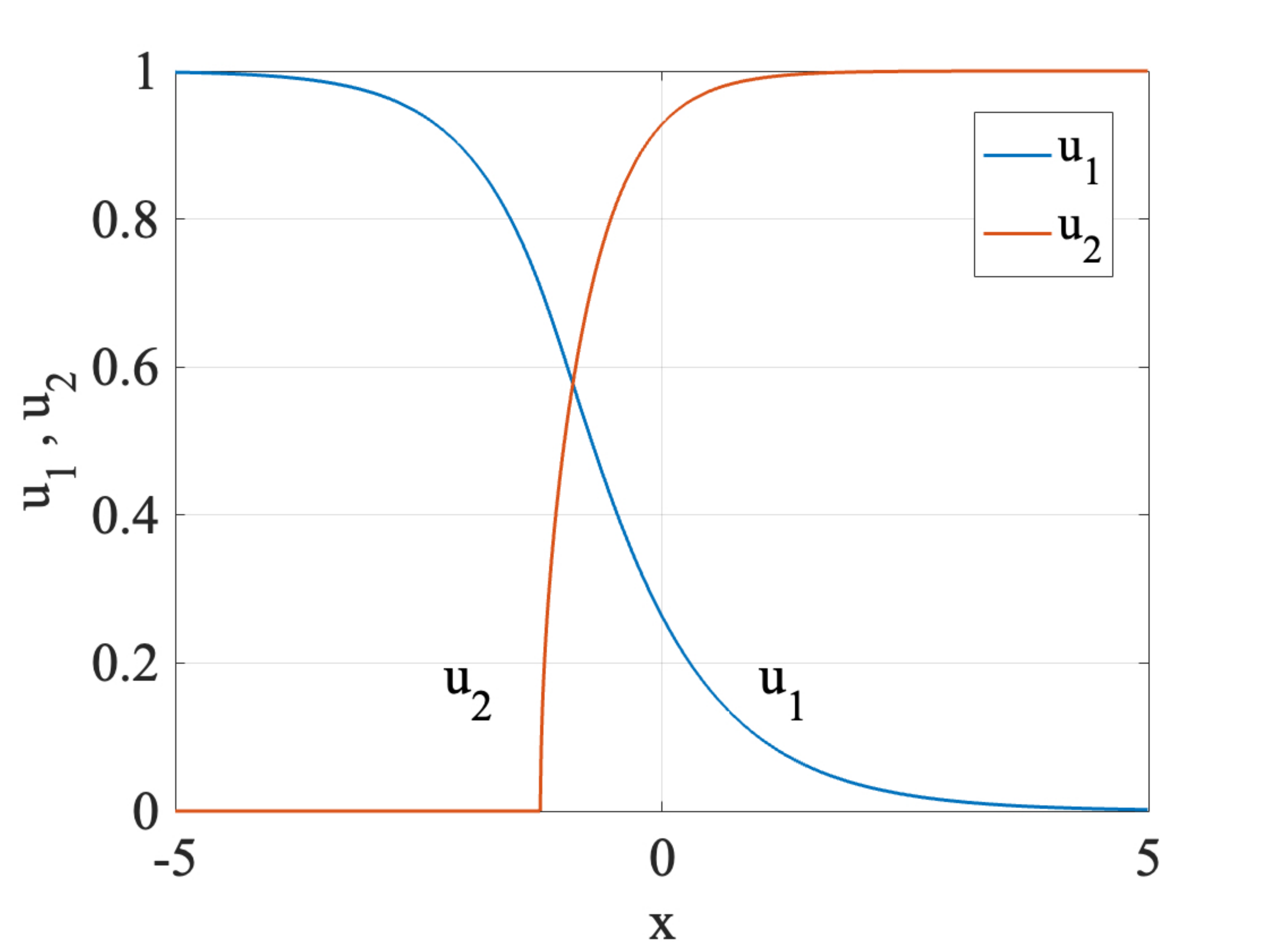}
\end{center}
\caption{An example of the asymmetric DW, as given by Eqs. (\protect\ref{x0}%
)-(\protect\ref{xi}), for $D=1$ and $G=2$. Note that the coordinate of the
stitch point is, in this case, $x_{0}\approx -1.25$, as per Eq. (\protect\ref%
{x0}), and the \textquotedblleft virtual center" of the bright-soliton
segment of $u_{1}(x)$ is located at $\protect\xi \approx -1.80$, as per Eq. (%
\protect\ref{xi}).}
\label{fig1}
\end{figure}

It is easy to check that expression (\ref{r1exact}) satisfies continuity
demands for $u_{1}(x)$ and $du_{1}/dx$ at $x=x_{0}$, and expression (\ref%
{r2exact}) provides the continuity of $u_{2}(x)$ at the same point. The
continuity of $dx_{2}/dt$ at $x=x_{0}$ is not required, as Eq. (\ref{D=0})
does not include derivatives. It is worthy to note that, unlike the
above-mentioned exact symmetric solution (\ref{exact1}), which exists, as an
isolated one, solely at $G=3$, the asymmetric solution given by Eqs. (\ref%
{x0})-(\ref{xi}) exists for all values of $G>1$. A typical example of the
solution is displayed, for $D=1$ and $G=2$, in Fig. \ref{fig1}.

It is relevant to mention that a more complex type of asymmetric DWs was
considered, in a numerical form, in Ref. \cite{Rotstein}. It is a wall
between two uniform bimodal states (square-lattice patterns), built as per
Eq. (\ref{U}), one with a pair of wave vectors corresponding to angles $%
\left( \theta _{1}=0,\theta _{2}=\pi /2\right) $, and the other pair rotated
by $\pi /4$, i.e., with $\theta _{1,2}=\pm \pi /4$.

\section{The DW in the symmetric system with linear coupling, and the effect
of the trapping potential}

\subsection{The exact solution}

The system of Eqs. (\ref{1}) and (\ref{2}), as it appears in above-mentioned
realizations in optics and BEC, may also include linear mixing between the
components. In particular, this effect is produced by twist applied to a
bulk optical waveguide \cite{twist,twist2}. A similar effect in binary BEC,
\textit{viz}., mutual inter-conversion of two atomic states, which form the
binary condensate, may be induced by the resonant radio-frequency field \cite%
{radio}. The respectively modified symmetric system of Eqs. (\ref{1}) and (%
\ref{2}) is
\begin{eqnarray}
D\frac{d^{2}u_{1}}{dx^{2}}+u_{1}\left( 1-u_{1}^{2}-Gu_{2}^{2}\right)
+\lambda u_{2} &=&0,  \label{lambda1} \\
D\frac{d^{2}u_{2}}{dx^{2}}+u_{2}\left( 1-u_{2}^{2}-Gu_{1}^{2}\right)
+\lambda u_{1} &=&0,  \label{lambda2}
\end{eqnarray}%
where real $\lambda $ is the linear-coupling coefficient. In fact, Eqs. (\ref%
{lambda1}) and (\ref{lambda2}) apply to the RB convection too, in the case
when periodic corrugation of the bottom of the convection cell, with
amplitude $\sim \lambda $ and wave vector $\mathbf{k}_{1}+\mathbf{k}_{2}$
(see Eq. (\ref{U})), gives rise to the \textit{linear cross-gain}, which is
used in many laser setups that are similar to thermal convection \cite%
{cross-,cross-2}.

\begin{figure}[tbp]
\begin{center}
\includegraphics[width=0.60\textwidth]{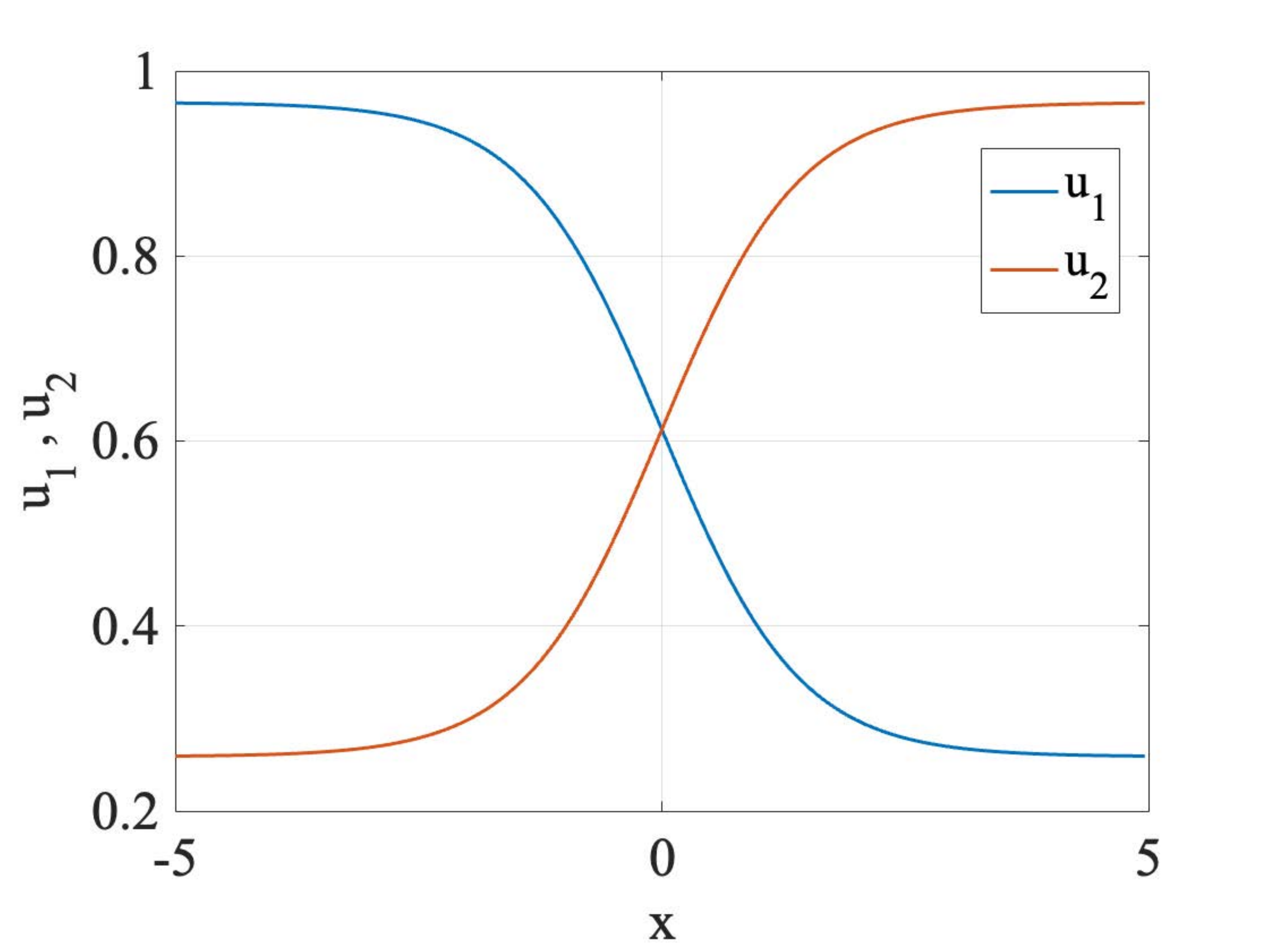}
\end{center}
\caption{An example of the symmetric DW in the presence of the linear
coupling, as given by Eqs. (\protect\ref{exact2}) and (\protect\ref{bc2}),
for $D=\protect\lambda =1/2$. Note that the asymptotic values of the
components at $x\rightarrow \pm \infty $, as given by Eqs. (\protect\ref{bc2}%
) are, in this case, $u_{1}(x\rightarrow -\infty )=u_{2}\left( x=+\infty
\right) \approx 0.97$ and $u_{1}(x\rightarrow +\infty )=u_{2}\left(
x=-\infty \right) \approx 0.26$.}
\label{fig2}
\end{figure}

The system of Eqs. (\ref{lambda1}) and (\ref{lambda2}) with $G=3$ admits an
exact DW solution, which is an extension of its counterpart (\ref{exact1}):%
\begin{equation}
\left\{
\begin{array}{c}
u_{1}(x) \\
u_{2}(x)%
\end{array}%
\right\} =\frac{1}{2}\left\{
\begin{array}{c}
\sqrt{1+\lambda }-\sqrt{1-\lambda }\tanh \left( \sqrt{\frac{1-\lambda }{2D}}%
x\right) \\
\sqrt{1+\lambda }+\sqrt{1-\lambda }\tanh \left( \sqrt{\frac{1-\lambda }{2D}}%
x\right)%
\end{array}%
\right\} .  \label{exact2}
\end{equation}%
Due to the action of the linear mixing, b.c. (\ref{bc}) are replaced by%
\begin{gather}
u_{1}\left( x\rightarrow -\infty \right) =u_{2}\left( x\rightarrow +\infty
\right) =\frac{1}{2}\left( \sqrt{1+\lambda }+\sqrt{1-\lambda }\right) ,
\notag \\
u_{1}\left( x\rightarrow +\infty \right) =u_{2}\left( x\rightarrow -\infty
\right) =\frac{1}{2}\left( \sqrt{1+\lambda }-\sqrt{1-\lambda }\right) .
\label{bc2}
\end{gather}%
These solutions exist for all values of $0\leq \lambda <1$. A typical
example is displayed in Fig. \ref{fig2}.

\subsection{Effects of the trapping potential}

The realization of the system in terms of the binary BEC should include,
generally speaking, the trapping HO potential, which is normally used in the
experiment \cite{Pit}. The accordingly modified system of Eqs. (\ref{lambda1}%
) and (\ref{lambda2}) is%
\begin{eqnarray}
D\frac{d^{2}r_{1}}{dx^{2}}+r_{1}\left( 1-r_{1}^{2}-Gr_{2}^{2}\right)
+\lambda r_{2} &=&\frac{\aleph ^{2}}{2}x^{2}r_{1},  \label{omega1} \\
D\frac{d^{2}r_{2}}{dx^{2}}+r_{2}\left( 1-r_{2}^{2}-Gr_{1}^{2}\right)
+\lambda r_{1} &=&\frac{\aleph ^{2}}{2}x^{2}r_{2},  \label{omega2}
\end{eqnarray}%
where $\aleph ^{2}$ is the strength of the OH potential. DW solutions of the
system of Eqs. (\ref{omega1}) and (\ref{omega2}) were addressed in Ref. \cite%
{Merh}. In the absence of the linear coupling ($\lambda =0$), a rigorous
mathematical framework for the analysis of such solutions was elaborated in
Ref. \cite{Peli}.

If the HO\ trap is weak, \textit{viz}., $\aleph ^{2}\ll 4/\left( 1-\lambda
\right) $, the DW solution trapped in the OH potential takes nearly constant
values, close to those in Eq. (\ref{bc2}), in the region of%
\begin{equation}
2D/\left( 1-\lambda \right) \ll x^{2}\ll 8D/\aleph ^{2}.  \label{intermed}
\end{equation}%
On the other hand, at $x^{2}\rightarrow \infty $ solutions generated by Eqs.
(\ref{omega1}) and (\ref{omega2}) decay similar to eigenfunctions of the HO
potential in quantum mechanics, \textit{viz}.,
\begin{eqnarray}
r_{1,2} &\approx &\varrho _{1,2}|x|^{\gamma }\exp \left( -\frac{\aleph }{2%
\sqrt{2D}}x^{2}\right) ,  \label{r12} \\
\gamma &=&\frac{1+\lambda }{\sqrt{2D}\aleph }-\frac{1}{2},  \label{gamma}
\end{eqnarray}%
where $\varrho _{1,2}$ are constants. In the case of $\lambda =0$, the
asymptotic tails (\ref{r12}) follow the structure of solution (\ref{exact1}),
i.e., $\varrho _{1}\left( x\rightarrow +\infty \right) =\varrho _{2}\left(
x\rightarrow -\infty \right) =0$ and $\varrho _{1}\left( x\rightarrow
-\infty \right) =\varrho _{2}\left( x\rightarrow +\infty \right) \neq 0$. On
the other hand, the linear mixing, $\lambda \neq 0$, makes the tail
symmetric with respect to the two components, with $\varrho _{1}\left(
|x|\rightarrow \infty \right) =\varrho _{2}\left( |x|\rightarrow \infty
\right) \neq 0$. Note that $\gamma =0$ in Eq. (\ref{gamma}) with $\lambda =0$
is tantamount to the case when values of $\aleph $ and $D$ in Eqs. (\ref%
{omega1}) and (\ref{omega2}) correspond to the ground state of the HO
potential.

\section{DW-bright-soliton complexes}

\subsection{An exact solution for the composite state}

The DW formed by two immiscible PWs may serve as an effective potential for
trapping an additional PW mode. To address this possibility, it is relevant
to consider the symmetric configuration, with $D_{1}=D_{2}\equiv D$ (see Eq.
(\ref{D})), and wave vector $\mathbf{k}_{v}$ of the additional PW mode, $%
v(x) $, directed along the bisectrix of the angle between the DW-forming
wave vectors $\mathbf{k}_{1}$ and $\mathbf{k}_{2}$, i.e., along axis $x$
(hence Eq. (\ref{D}) yields $D_{v}=1$). The corresponding system of three
coupled stationary real GL equations is%
\begin{eqnarray}
D\frac{d^{2}u_{1}}{dx^{2}}+u_{1}\left( 1-u_{1}^{2}-Gu_{2}^{2}-gv^{2}\right)
&=&0,  \label{u1g} \\
D\frac{d^{2}u_{2}}{dx^{2}}+u_{2}\left( 1-u_{2}^{2}-Gu_{1}^{2}-gv^{2}\right)
&=&0,  \label{u2g}
\end{eqnarray}%
\begin{equation}
\frac{d^{2}v}{dx^{2}}+\left( 1-v^{3}-g\left( u_{1}^{2}+u_{2}^{2}\right)
\right) v=0,  \label{vv}
\end{equation}%
where $g>0$ is the constant of the nonlinear interaction between components $%
u_{1,2}$ and $v$.

The system of Eqs. (\ref{u1g})-(\ref{vv}) admits the following exact
solution, in the form of the DW of components $u_{1,2}(x)$ coupled to a
bright-soliton profile of $v(x)$:%
\begin{equation}
\left\{
\begin{array}{c}
u_{1}(x) \\
u_{2}(x)%
\end{array}%
\right\} =\frac{1}{2}\left\{
\begin{array}{c}
1-\tanh \left( \sqrt{g-1}x\right) \\
1+\tanh \left( \sqrt{g-1}x\right)%
\end{array}%
\right\} ,  \label{exact-v}
\end{equation}%
\begin{equation}
v(x)=\sqrt{2-\frac{3}{2}g}\mathrm{sech}\left( \sqrt{g-1}x\right) .
\label{exact-vv}
\end{equation}%
This solution is valid under the condition that coefficients $G$ and $D$ in
Eqs. (\ref{u1g}) and (\ref{u2g}) take the following particular values,%
\begin{equation}
G=3-8g+6g^{2},  \label{G}
\end{equation}%
\begin{equation}
D=\frac{1}{2}\left( 3g-1\right) .  \label{DD}
\end{equation}%
As is follows from Eq. (\ref{exact-vv}), $g$ is a free parameter of this
solution, which may take values in a narrow interval,
\begin{equation}
1<g<4/3  \label{4/3}
\end{equation}%
(see also Eq. (\ref{1-2}) below). According to Eqs. (\ref{G}) and (\ref{DD}%
), the interval (\ref{4/3}) of the variation of $g$ corresponds to
coefficients $G$ and $D$\ varying in intervals%
\begin{equation}
1<G<3;~1<D<3/2.  \label{GD}
\end{equation}%
Thus, adding the $v$ component lifts the degeneracy of the exact DW solution
(\ref{exact1}), which exists solely at $G=3$.

Recall that, in the model of convection patterns, coefficient $D$, as given
by Eq. (\ref{D}), cannot take values $D>1$, which disagrees with Eq. (\ref%
{GD}). However, values $D>1$ are relevant for systems of GP equations for
the heteronuclear three-component BEC. In the latter case, $D$ is the ratio
of atomic masses of the different species which form the triple immiscible
BEC. Similarly, $D$ is the ratio of values of the normal group-velocity
dispersion of copropagating waves in the temporal-domain realization of the
real GL equations in nonlinear fiber optics \cite{optical-DW}.

\begin{figure}[tbp]
\begin{center}
\includegraphics[width=0.60\textwidth]{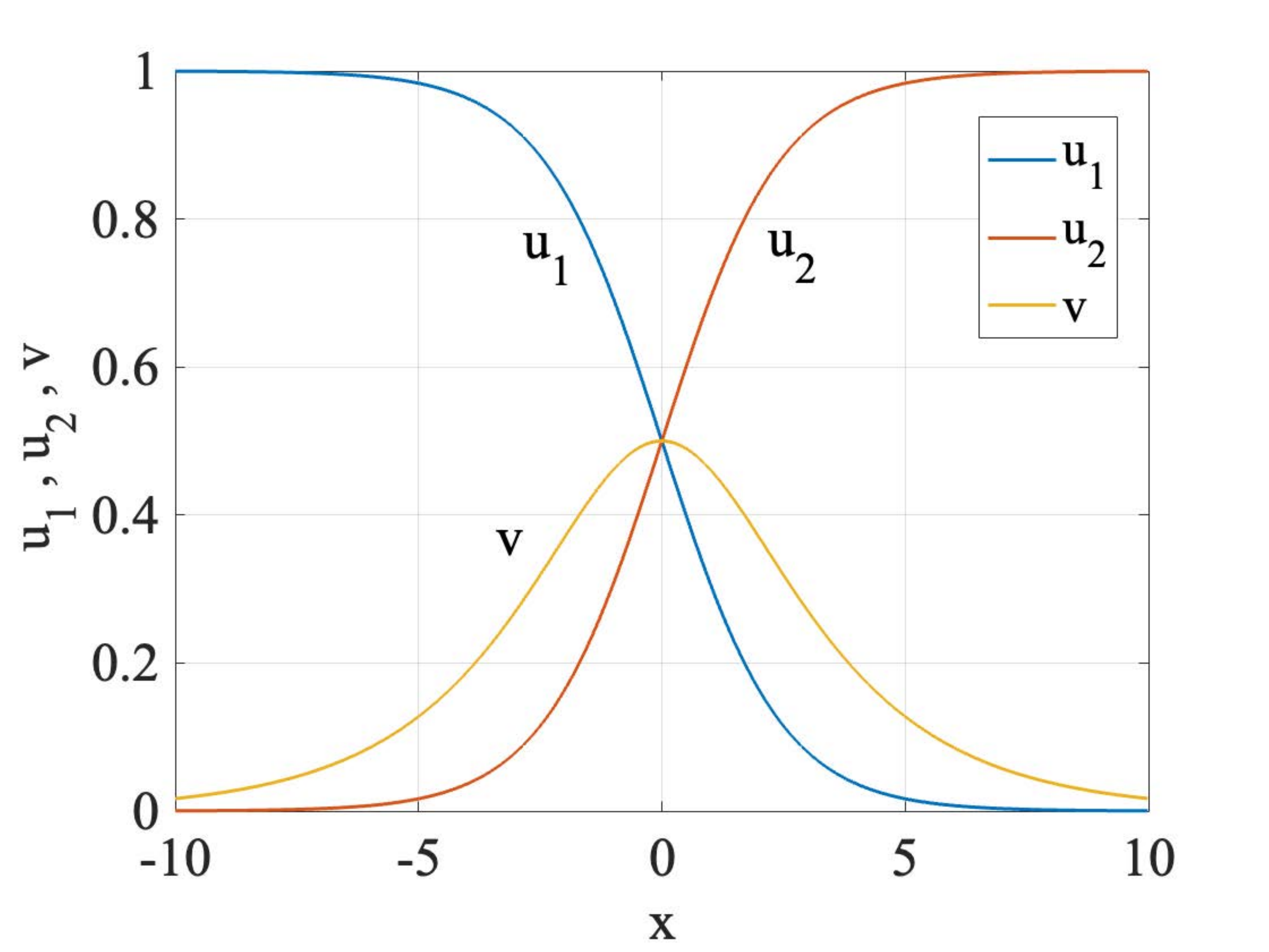}
\end{center}
\caption{An example of the exact solution for the DW-bright-soliton complex,
given by Eqs. (\protect\ref{exact-v}) and (\protect\ref{exact-vv}), for $%
g=7/6,G=11/6,$ and $D=5/4$.}
\label{fig3}
\end{figure}

An example of the DW-bright-soliton complex is displayed in Fig. \ref{fig3}
for $g=7/6$, in which case Eqs. (\ref{G}) and (\ref{DD}) yield $G=11/6$ and $%
D=5/4$ (according to Eqs. (\ref{G}) and (\ref{D})). The fact that the
respective soliton's amplitude, which is $\sqrt{2-3g/2}=1/2$ according to
Eq. (\ref{exact-vv}), coincides with the mid value of the DW components (\ref%
{exact-v}), is a peculiarity of this particular case.

\subsection{The bifurcation of the creation of the composite state in the
general case}

If relation (\ref{G}) is not imposed on the interaction coefficients $g$ and
$G$, the solution for the composite state cannot be found in an exact form.
Nevertheless, it is possible to identify \textit{bifurcation points} at
which component $v$ with an infinitesimal amplitude appears. To this end,
Eq. (\ref{vv}) should be used in the form linearized with respect to $v$:
\begin{equation}
\frac{d^{2}v}{dx^{2}}+\left\{ 1-g\left[ u_{1}^{2}(x)+u_{2}^{2}(x)\right]
\right\} v=0.  \label{rho}
\end{equation}%
This linear equation can be exactly solved for $u_{1}(x)=u_{2}(x)$ given by
expression (\ref{exact1}), in the case of $G=3$, while parameters $D$ and $g$
may take arbitrary values. Indeed, using the commonly known results for the P%
\"{o}schl-Teller potential in quantum mechanics, it is easy to find that Eq.
(\ref{rho}) with the effective potential corresponding to Eq. (\ref{exact1})
gives rise to spatially even eigenmodes in the form of%
\begin{equation}
v(x)=\mathrm{const}\cdot \left[ \mathrm{sech}\left( x/\sqrt{2D}\right) %
\right] ^{\alpha },  \label{eigen}
\end{equation}%
at a special value of the interaction coefficient, which identifies the
bifurcation producing the composite state:%
\begin{equation}
g_{\mathrm{bif}}=D^{-1}\left( 1+2D\mp \sqrt{1+2D}\right) ,  \label{bif}
\end{equation}%
the respective value of power $\alpha $ in expression (\ref{eigen}) being%
\begin{equation}
\alpha =\sqrt{2\left( 1+D\mp \sqrt{1+2D}\right) }.  \label{alpha}
\end{equation}%
The values given by Eqs. (\ref{bif}) and (\ref{alpha}) with the top sign
from $\mp $ correspond to the bifurcation creating a fundamental composite
state (the ground state, in terms of the quantum-mechanical analog) at $g>g_{%
\mathrm{bif}}$, while the bottom sign represents a higher-order bifurcation
(alias the second excited state, in the language of quantum mechanics; the
first excited state, is a spatially odd mode which is considered below).
While it is obvious that the fundamental bifurcation creates a stable
composite state, the ones produced by higher-order bifurcations may be
unstable.

Further, varying coefficient $D$ of the modes forming the underlying DW
between $D=0$ and $D=\infty $ (recall that the convection model corresponds
to $D<1,$ while the realizations in optics and BEC admit $D>1$), Eq. (\ref%
{bif}) demonstrates monotonous variation of the bifurcation point in
interval
\begin{equation}
g_{\mathrm{bif}}(D=0)\equiv 1<g_{\mathrm{bif}}<2\equiv g_{\mathrm{bif}%
}\left( D\rightarrow \infty \right) .  \label{1-2}
\end{equation}%
It extends interval (\ref{4/3}) in which exact composite states with a
finite amplitude were found above, see Eqs. (\ref{exact-v})-(\ref{DD}).

An odd linear mode produced by Eq. (\ref{rho}) with $u_{1,2}(x)$ taken from
Eq. (\ref{exact1}) is looked for as%
\begin{equation}
v(x)=\mathrm{const}\cdot \sinh \left( x/\sqrt{2D}\right) \left[ \mathrm{sech}%
\left( x/\sqrt{2D}\right) \right] ^{\beta }.  \label{beta}
\end{equation}%
The corresponding exact solution for the quantum-mechanical P\"{o}%
schl-Teller potential has
\begin{eqnarray}
\beta  &=&\sqrt{2D\left( g_{\mathrm{bif}}^{(\beta )}-1\right) }+1\equiv 1+%
\sqrt{2\left( D+7-3\sqrt{2D+5}\right) },  \label{beta-solution} \\
g_{\mathrm{bif}}^{(\mathrm{odd})} &=&D^{-1}\left( 2D+7-3\sqrt{2D+5}\right) .
\label{beta-bif}
\end{eqnarray}%
This solution is valid for $D>2$. As $D$ varies from $2$ to $\infty $,
expression (\ref{beta-bif}) monotonously increases from $g_{\mathrm{bif}}^{(%
\mathrm{odd})}=1$ to $g_{\mathrm{bif}}^{(\mathrm{odd})}=2$. Note that, at $%
D=2$, Eq. (\ref{bif}) yields $g_{\mathrm{bif}}(D=2)=(1/2)(5-\sqrt{5})\approx
\allowbreak 1.382$. Actually, at all values of $D\geq 2$, the value of $g_{%
\mathrm{bif}}^{(\mathrm{odd})}$ is \emph{smaller} than $g_{\mathrm{bif}}$,
which is given by Eq. (\ref{bif}) for the fundamental (even) mode. This fact
implies that, with the increase of $g$, the bifurcation creating the
spatially odd mode in the $v$ component happens \emph{earlier} than the
bifurcation which creates the even mode.

At $D=2$, Eq. (\ref{beta-solution}) yields $\beta =1$, which corresponds to
the delocalized eigenmode (\ref{beta}), $v(x)=\mathrm{const}\cdot \tanh
\left( x/\sqrt{2D}\right) $. With the increase of $D$, $\beta $ increases
monotonously towards $\beta \rightarrow \infty $.

\section{Domain walls between traveling waves}

\subsection{The sink and source in the two-component system}

In the simplest case, the system of GL equations for counterpropagating
dissipative waves can be written in the form which neglects dispersive
effects but includes the opposite group velocities, $\pm c$ \cite{traveling}%
:
\begin{eqnarray}
\frac{\partial u_{1}}{\partial t}+c\frac{\partial u_{1}}{\partial x}
&=&u_{1}+\frac{\partial ^{2}u_{1}}{\partial x^{2}}-u_{1}\left(
|u_{1}|^{2}+G|u_{2}|^{2}\right) ,  \label{u1} \\
\frac{\partial u_{2}}{\partial t}-c\frac{\partial u_{1}}{\partial x}
&=&u_{2}+\frac{\partial ^{2}u_{2}}{\partial x^{2}}-u_{2}\left(
|u_{2}|^{2}+G|u_{1}|^{2}\right) .  \label{u2}
\end{eqnarray}%
These equations, unlike Eqs. (\ref{1d/dt}) and (\ref{2d/dt}), do not admit
the gradient representation. Nevertheless, the stationary form of Eqs. (\ref%
{u1}) and (\ref{u2}) amounts to real equations:
\begin{eqnarray}
+c\frac{du_{1}}{dx} &=&\frac{d^{2}u_{1}}{dx^{2}}+u_{1}\left(
1-u_{1}^{2}-Gu_{2}^{2}\right) ,  \label{r11} \\
-c\frac{du_{2}}{dx} &=&\frac{d^{2}u_{2}}{dx^{2}}+u_{2}\left(
1-u_{2}^{2}-Gu_{1}^{2}\right) .  \label{r22}
\end{eqnarray}%
In this case, the relevant b.c. keeps the form of Eq. (\ref{bc}).

\begin{figure}[tbp]
\begin{center}
\subfigure[]{\includegraphics[width=0.44\textwidth]{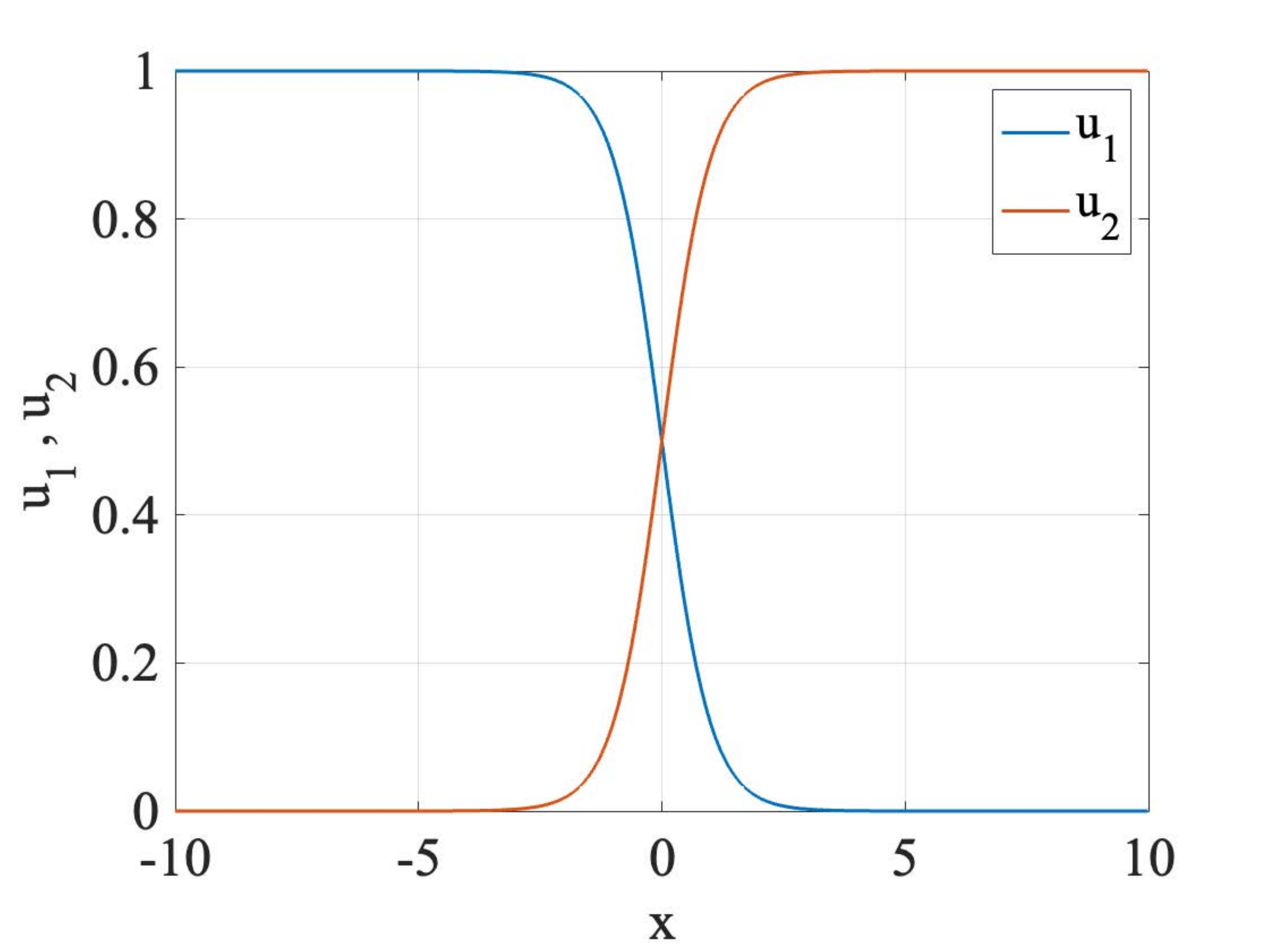}}%
\subfigure[]{\includegraphics[width=0.44\textwidth]{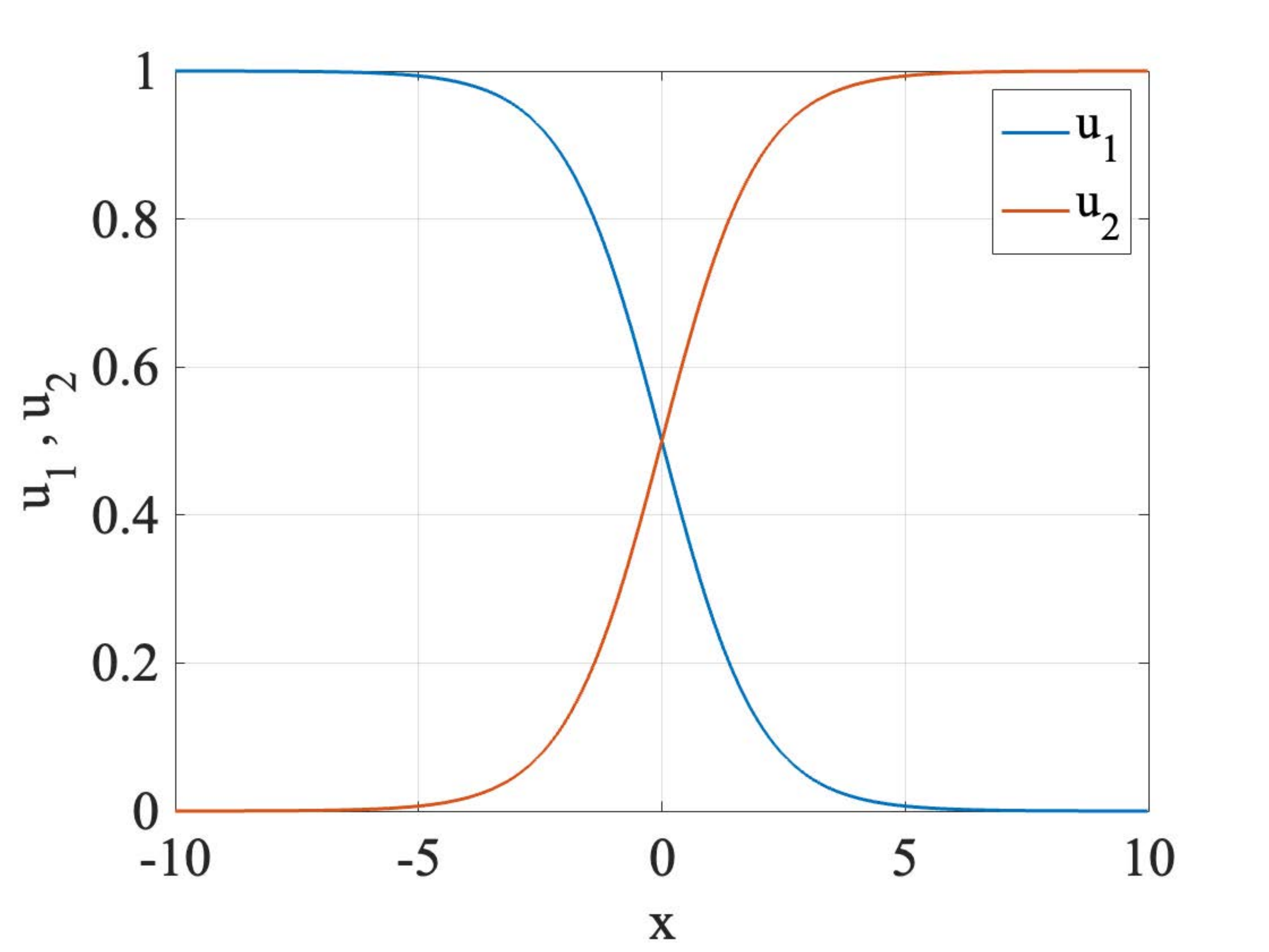}}
\end{center}
\caption{Examples of the exact stationary solution for coupled traveling
waves, given by Eq. (\protect\ref{exact3}): (a) a \textit{sink}, for $G=7$
and $c=+1$ in Eqs. (\protect\ref{r11}) and (\protect\ref{r22}); (b) a
\textit{source}, for $G=1$ and $c=-1$.}
\label{fig4}
\end{figure}

An exact solution to Eqs. (\ref{r11}) and (\ref{r22}) can be found following
the pattern of Eq. (\ref{exact1}):%
\begin{equation}
\left\{
\begin{array}{c}
u_{1}(x) \\
u_{2}(x)%
\end{array}%
\right\} =\frac{1}{2}\left\{
\begin{array}{c}
1-\tanh \left( \left( \sqrt{8+c^{2}}+c\right) (x/4\right) \\
1+\tanh \left( \left( \sqrt{8+c^{2}}+c\right) (x/4\right)%
\end{array}%
\right\} ,  \label{exact3}
\end{equation}%
in the case when the cross-interaction coefficient takes a specific value
\begin{equation}
G-3=c\left( \sqrt{8+c^{2}}+c\right) ,  \label{G-3}
\end{equation}%
or, inversely,
\begin{equation}
c=\left( G-3\right) /\sqrt{2\left( G+1\right) }.  \label{c}
\end{equation}%
Thus, this solution lifts the degeneracy of the \textquotedblleft old" one (%
\ref{exact1}), which exists solely at $G=3$. Further, it follows from Eq. (%
\ref{G-3}) and (\ref{vv}) that $\mathrm{sgn}(v)=\mathrm{sgn}\left(
G-3\right) $, hence, taking into regard b.c. (\ref{bc}), one concludes that
the exact solution (\ref{exact3}) represents a \textit{sink} of traveling
waves ($c>0$) for $G>3$, and a \textit{source} ($c<0$) for $G<3$. Typical
examples of the sink and source are displayed in Figs. \ref{fig4}(a) and
(b), respectively. The solution of the latter type exists even in the case
of $G<1$, when the two components are miscible; in that case, the separation
between them in the DW pattern is maintained by the opposite group
velocities, which pull the components apart, preventing the onset of the
mixing. In fact, it follows from Eq. (\ref{c}) that the solution persists
even in the range of moderately strong \textit{attraction} between the
component, $-1<G<0$. Note that the pressure of the incoming stationary flows
makes the sink mode in Fig. \ref{fig4}(a) conspicuously narrower than its
source counterpart drawn in Fig. \ref{fig4}(b) for the same absolute value
of the group velocities, $|c|=1$. The source is broader as it is stretched
by the egressing flows, even if it is plotted for much weaker mutual
repulsion between the components ($G=1$) than the sink, which pertains to $%
G=7$.

\subsection{The composite state in the three-component system}

The pair of counterpropagating traveling waves which can trap the additional
standing one are described by the following generalization of Eqs. (\ref{r11}%
) and (\ref{r22}):
\begin{eqnarray}
+c\frac{du_{1}}{dx} &=&\frac{d^{2}u_{1}}{dx^{2}}+u_{1}\left(
1-u_{1}^{2}-Gu_{2}^{2}-gv^{2}\right) ,  \label{ext1} \\
-c\frac{du_{2}}{dx} &=&\frac{d^{2}u_{2}}{dx^{2}}+u_{2}\left(
1-u_{2}^{2}-Gu_{1}^{2}-gv^{2}\right) ,  \label{ext2}
\end{eqnarray}%
to which an equation for the standing mode is added, cf. Eq. (\ref{vv}):
\begin{equation}
\frac{d^{2}v}{dx^{2}}+\left( 1-v^{2}-g\left( u_{1}^{2}+u_{2}^{2}\right)
\right) v=0,  \label{ext3}
\end{equation}%
An exact solution of Eqs. (\ref{ext1})-(\ref{ext3}) can be found for free
parameters $g$ and $c$:%
\begin{eqnarray}
u_{1,2}(x) &=&\frac{1}{2}\left( 1\mp \tanh \left( \sqrt{g-1}x\right) \right)
,  \label{ext4} \\
v(x) &=&\sqrt{2-\frac{3}{2}g}\mathrm{sech}\left( \sqrt{g-1}x\right) ,
\label{ext5}
\end{eqnarray}%
\begin{eqnarray}
G-3 &=&2g\left( 3g-4\right) +4c\sqrt{g-1},  \label{ext6} \\
D &=&\frac{c}{2\sqrt{g-1}}+\frac{1}{2}\left( 3g-1\right) ,  \label{ext7}
\end{eqnarray}%
cf. Eqs. (\ref{exact-v})-(\ref{DD}). As it is seen from Eq. (\ref{ext6}),
the interaction with the soliton-shaped standing wave shifts the boundary
between the sink and source of the traveling waves from the above-mentioned
point, $G=3$.

Further, if, in the absence of $v(x)$, the bimodal solution for traveling
waves is given by Eqs. (\ref{exact3})-(\ref{c}), the consideration of the
bifurcation which gives rise to infinitesimal even and odd modes in the $v$
component produces the same results as given above, respectively, by Eqs. (%
\ref{eigen})-(\ref{alpha}) and (\ref{beta})-(\ref{beta-bif}), with $D$
replaced by%
\begin{equation}
D_{\mathrm{eff}}=\frac{8D^{2}}{\left( c+\sqrt{c^{2}+8D}\right) ^{2}}
\label{Deff}
\end{equation}%
(note that, in the limit of $D\rightarrow \infty $, Eq. (\ref{Deff}) yields $%
D_{\mathrm{eff}}\approx D$). In particular, the value of $D=2$, at which the
odd modes appears above, is replaced by $D_{\mathrm{eff}}=2$, which
corresponds to $D=2+c$.

\section{Conclusion}

The aim of this paper is to report new exact solutions for the well-known
problem of constructing DW (domain-wall) solutions of the system of coupled
real GL (Ginzburg-Landau) equations. These equations apply to modeling DW
patterns (alias grain boundaries) in RB (Rayleigh-B\'{e}nard) convection,
nonlinear optics, and binary BEC. Even if exact solutions cannot be generic
ones, particular analytical solutions are quite useful, as they provide
direct insight into the structure of DW states. A particular exact solution
for the symmetric DW was found long ago in Ref. \cite{Trib-DW}. It is an
isolated solution, which exists at the single value of the cross-interaction
coefficient, $G=3$. In this work, first, an exact solution for strongly
asymmetric DWs is found in the form of Eqs. (\ref{x0})-(\ref{xi}), for the
system in which the diffusion term is present in one component only. Unlike
the \textquotedblleft old" exact solution for the symmetric DW, the newly
found one is available at all values of $G>1$, which is the fundamental
condition for immiscibility of the two components. An exact solution for the
symmetric DW, in the system including the linear coupling between the
components, is found too, given by Eq. (\ref{exact2}). In addition to that,
the effect of the trapping harmonic-oscillator potential on the DW is
considered, leading to the asymptotic form of the solution presented by Eqs.
(\ref{r12}) and (\ref{gamma}). Another essential finding is exact solution (%
\ref{u1g})-(\ref{vv}) for the system of three coupled GL equations for a
composite state built of a symmetric DW between two components and a bright
soliton in the third one. This solution also lifts the degeneracy of the
\textquotedblleft old" one, fixed by $G=3$. In addition to this result, the
location of the bifurcations, which create the composite states from the
two-component DW, are found in the exact form, as given by Eqs. (\ref{eigen}%
)-(\ref{alpha}) or (\ref{beta})-(\ref{beta-bif}) for the bifurcations
creating, respectively, the spatially even (fundamental) or odd component in
the third component. The stability of all these exact solutions is provided
by the gradient structure of the underlying systems of time-dependent GL
equations. Finally, another exact stationary solution, provided by Eqs. (\ref%
{exact3})-(\ref{c}), is generated by the system of GL equations governing
the interaction of counterpropagating waves with opposite group velocities.
The solution also lifts the degeneracy condition $G=3$ and, depending on the
sign of $G-3$, it represents either a sink or source of the waves. The
source-type states exists even in the case of $G<1$, when the immiscibility
condition does not hold for the interacting components. The latter solution
is complemented by the exact composite one, given by Eqs. (\ref{ext4})-(\ref%
{ext7}), which includes the localized mode in the third (standing)
component. The respective bifurcations are identified too, by means of Eq. (%
\ref{Deff}).

As an extension of this work, it may be relevant to develop the analysis for
families of generic DW states originating from the particular exact
solutions reported in this paper. This can be done by means of the
perturbation theory and numerical methods.

\section*{Acknowledgments}

I thank Michael Tribelsky, Alexander Nepomnyashchy, and Dmitry Pelinovsky
for valuable discussions. The help of Zhaopin Chen in producing plots
included in this paper is highly appreciated. This work was supported, in
part, by Israel Science Foundation through grant No. 1286/17.

\end{document}